\documentclass[aps,prb,twocolumn,10pt,superscriptaddress]{revtex4-2}
\usepackage{xcolor}
\usepackage{graphicx}
\usepackage{amssymb,amsmath,bm}
\usepackage{braket}
\usepackage{epstopdf}
\usepackage{appendix}
\begin{document}

\title{Rashba-Cavity Engineering of Energy Gaps and Spin-Hall Conductivity \\ in Semiconducting Artificial Graphene}

\author{Maryam Mansouri}
\affiliation{Department of Condensed Matter Physics, Yerevan State University, Yerevan, Armenia}

\author{Vram Mughnetsyan}
\affiliation{Department of Condensed Matter Physics, Yerevan State University, Yerevan, Armenia}

\author{Armen Harutyunyan}
\affiliation{Department of Condensed Matter Physics, Yerevan State University, Yerevan, Armenia}

\author{Albert Kirakosyan}
\affiliation{Department of Condensed Matter Physics, Yerevan State University, Yerevan, Armenia}

\author{Vidar Gudmundsson}
\affiliation{Science Institute, University of Iceland, Dunhaga 3, IS-107 Reykjavik, Iceland}

\date{\today}

\begin{abstract}
We investigate the combined effect of a far - infrared cavity field and Rashba spin - orbit interaction on the band structure and transport properties of artificial graphene composed of quasi-2D InAs/GaAs quantum dots. The coupling to cavity photons is modeled by constructing a complete basis as the tensor product of the electronic Hilbert space and the Fock space. Our numerical calculations for the system embedded in a linear cavity predict the existence of both type-I and type-II Dirac points which can be distinguished by their response to Rashba interaction. Namely, Rashba coupling opens a gap at type-II Dirac points, while type-I Dirac points remain gapless. The possibility of gap-opening for type-II Dirac points is demonstrated analytically as well. For both cylindrical and linear cavities, we demonstrate the formation of electron - photon hybrid states and Rabi splittings between energy minibands. Multiple crossings, and anticrossings between Dirac-band replicas produce pronounced modifications of the spin-Hall conductivity, including strong anisotropy and oscillatory behavior controlled by cavity geometry and polarization of photons. Our results show that the interplay between Rashba and cavity couplings governs Dirac-point physics and provides a route toward tunable polaritonic spin-transport in engineered nanostructures.

\end{abstract}

\maketitle

\section{Introduction}

The development of nanotechnology has progressed steadily for several decades and has accelerated significantly nowadays. The emergence of two-dimensional (2D) materials has provided a wide variety of material platforms for both fundamental research and technological applications at the atomic scale. Intensive investigation of 2D systems is expected to lead to the design of hybrid materials with enhanced functionalities, potentially enabling the synthesis of novel structures and the development of devices that outperform conventional technologies in various application domains \cite{Kumbhakar2023}. Graphene and monolayer hexagonal boron nitride (hBN), which are structural analogs, are prominent examples of advanced 2D materials. They have attracted considerable attention due to their exceptional mechanical, optical, and electronic properties \cite{Paine1990}. In particular, due to the honeycomb symmetry of the system quasiparticles in graphene are described by the rather exotic 2D massless Dirac Hamiltonian and thus offer fundamental insight into pseudorelativistic phenomena such as the iconic Klein paradox \cite{RevModPhys.81.109, katsnelson}. Linear energy dispersion in graphene allows tabletop realization of quantum relativistic phenomena that are otherwise inaccessible in high-energy physics experiments \cite{Geim2007}. Significant progress has also been achieved in the synthesis of ultra-thin hBN layers, suspended or supported on substrates such as molten gold, bulk copper, and thin copper films \cite{Lee2018,Chen2020}.

Artificial honeycomb lattices, particularly artificial graphene (AG), provide a versatile platform for studying and engineering systems hosting massless Dirac fermions, topological phases, and correlated electronic states \cite{Gomes2012,Tarruell2012,Singha2011}. One of the main motivations for investigating AG is the ability to access physical regimes that are impossible to realize in natural graphene, such as tunable lattice constants and controlled defect, strain, and edge engineering \cite{Ferrari2015}. Moreover, the tunability of Fermi velocity \cite{Downing2017} and the effective spin–orbit interaction (SOI) \cite{Mughnetsyan2019PRB} in AG make these systems attractive for the design of functional electronic and spintronic devices with controllable logical states \cite{Yung2013}.

Far-infrared (FIR) spectroscopy has proven to be an indispensable tool for investigating carrier excitations in the conduction band of two-dimensional electron systems.
The exceptionally high polarizability and mobility of a two-dimensional electron gas (2DEG) in a GaAs heterostructure make it an ideal experimental platform for achieving non-perturbative coupling between electrons and FIR cavity photons \cite{Zhang2016}.

In a modulated 2DEG placed inside a high-quality terahertz or FIR cavity the resulting quasiparticles take the form of miniband polaritons. Nonrelativistic quantum electrodynamics, often restricted to a single photon mode, together with various approaches to electron dynamics, has been employed to describe such electron–photon systems. For few-electron nanoscale systems, a range of toy models has been explored, with a primary focus on the electron–photon interaction \cite{Sun, Yuan2017}. 
In addition, frameworks that incorporate both para- and diamagnetic electron–photon couplings, along with Coulomb electron–electron interactions within a numerically exact diagonalization scheme, have been used to investigate both closed \cite{Jonasson2012, Gudmundsson2016} and open \cite{Jonsson2017, Gudmundsson2018} systems.

Since the existence of conventional or type-I Dirac points (DP) in graphene is intrinsically linked to the honeycomb lattice, the fundamental properties of massless Dirac quasiparticles are exceptionally robust and consequently resistant to manipulation. In contrast, artificial graphene platforms allow for precise lattice engineering, providing access to regimes of Dirac physics that are difficult, if not impossible, to realize in natural graphene \cite{Tarruell2012,Bellec2013,Rechtsman2013a,Rechtsman2013b,Xiang2018}.
Among the various approaches, strain engineering has emerged as a paradigmatic example, demonstrating that lattice anisotropy can modify the position of type-I DPs in AG \cite{Mughnetsyan2019PRB,Mughnetsyan2019SM} and drive the merging and annihilation of those in a natural graphene \cite{Pereira2009, Montambaux2009}, while aperiodicity can generate large pseudomagnetic fields \cite{Guinea2009}.
The discovery of type-II Dirac/Weyl semimetals \cite{Soluyanov2015, Deng2016, Huang2016a, Huang2016b, Yan2017}, with critically tilted cones and open isofrequency contours, has spurred studies of their electromagnetic analogs \cite{Xiao2016,Chen2016,Noh2017,Yang2017,Weight} and 2D variants \cite{Lin2017,Pyrialakos2017}, although their properties require lattice engineering. This search for exotic quasiparticles has recently extended to polaritonics \cite{Jacqmin2014,Nalitov2015,Karzig2015,Bardyn2015,Yi2016,Yuen-Zhou2016,Mann2018}, where the hybrid light–matter nature provides additional tunability, including the realization of topological polaritons \cite{Karzig2015,Bardyn2015}.

In this work, we investigate the combined effects of a FIR cavity field and Rashba SOI on the band structure and transport properties of AG composed of a honeycomb array of high-quality, high-purity InGaAs/GaAs quasi-2D quantum dots (QD). The coupling to cavity photons is modeled by constructing a complete basis as the tensor product of the electronic Hilbert space and the photon Fock space \cite{Malave2022,Gudmundsson2025}. Previous studies revealed a threefold splitting of type-I DPs in both natural graphene \cite{Zarea2009} and AG \cite{Mughnetsyan2019PRB}, while the original DP positions remain either unchanged or shifted from the Brillouin zone edge due to the strain anisotropy \cite{Mughnetsyan2019PRB}. Notably, neither Rashba SOI nor internal strain opens a gap between the minibands of AG. Our results demonstrate multiple crossings and anticrossings  of minibands induced by coupling to cavity modes and leading to pronounced changes in the system conductivity. Furthermore, we reveal the emergence of type-II DPs due to interaction with linearly polarized cavity modes. Interestingly, these additional type-II DPs can be effectively manipulated by means of Rashba SOI creating energy gaps. The above mentioned topological modifications of the energy minibands leave a clear signature in the spin-transport properties of polaritons, producing strongly anisotropic and field-controllable conductivity.

The paper is organized as follows: in Sec.\ II, we present a brief theoretical framework. The results are discussed in Sec.\ III, followed by the conclusions in Sec.\ IV. Acknowledgments are given in Sec. V.

\section{Theory}
\begin{figure}
\centering
\includegraphics[width=1\linewidth]{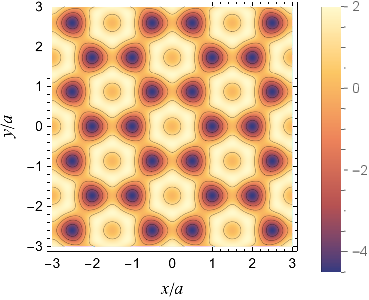}
\caption{The potential profile $V_{h}/V_{0}$ of the AG.}
\label{PotProf}
\end{figure}
\begin{figure*}
\centering
\includegraphics[width=1\linewidth]{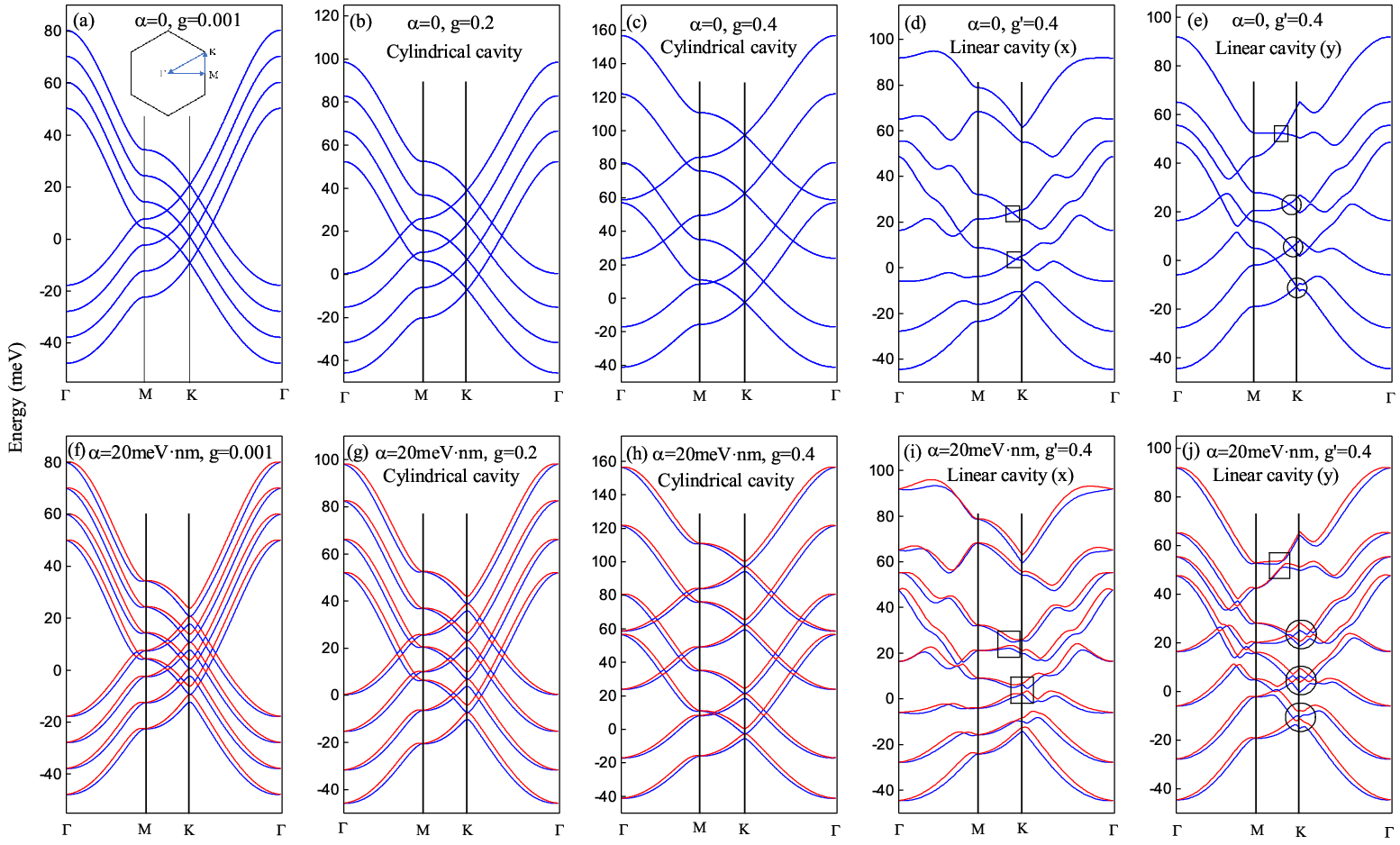}
\caption{Energy miniband dispersions along the chosen $k$-space path for the values of Rashba parameter $\alpha = 0$, and $20$ meV$\cdot$nm, and for interaction constants $g,g' = 0.001$ and  $0.4$ for cylindrical and linear cavities, respectively.}
\label{Bandstructure}
\end{figure*}
We consider a smoothly periodic potential (see Fig.\ \ref{PotProf}) to model an AG composed of quasi-2D QDs in the following analytical form
\begin{equation}
V_{\mathrm{h}}(\mathbf{r}) = V_0 \sum_{i=1}^{3} \big[ \cos(\mathbf{b}_i \cdot \mathbf{r}) - \cos(\mathbf{b}'_i \cdot \mathbf{r}) \big],
\label{Vh}
\end{equation}
with the vectors in the reciprocal space for two triangular lattices with periods $a$ and $\sqrt{3}a$ rotated by $\pi/2$ with regard to each other
\begin{equation}
\mathbf{b}_1 = \frac{4\pi}{\sqrt{3}a}
\{\tfrac{\sqrt{3}}{2}, \tfrac{1}{2}\},
\mathbf{b}_2 = \frac{4\pi}{\sqrt{3}a} \{\tfrac{\sqrt{3}}{2}, -\tfrac{1}{2}\},
\mathbf{b}_3 = \frac{4\pi}{\sqrt{3}a}\{0,1\}, \nonumber
\end{equation}
and
\begin{equation}
\mathbf{b}_1' = \frac{4\pi}{3a} \{-\tfrac{1}{2}, \tfrac{\sqrt{3}}{2}\},
\mathbf{b}_2' = \frac{4\pi}{3a} \{\tfrac{1}{2}, \tfrac{\sqrt{3}}{2}\},
\mathbf{b}_3' = \frac{4\pi}{3a} \{1,0\}, \nonumber
\end{equation}
where $a$ is the distance between the cites of the considered system. As is well known in the arrays of semiconducting QDs smooth profiles for the potential can be achieved, for instance, by inducing an interdiffusion between the compound materials of the heterostructure \cite{AZIZAGHCHEGALA2010,AZIZAGHCHEGALA2015}.

The total Hamiltonian of the system is as follows
\begin{equation}
    \mathcal{H} =H_\mathrm{0}+V_\mathrm{h}+H_{\mathrm{R}}+H_{\mathrm{e-ph}}+H_{\mathrm{ph}},
\label{Htot}
\end{equation}
where $H_0=\mathbf{p}^2/2m^{*}$, $\mathbf{p} =-i\hbar\bm{\nabla}$ is the momentum operator and $m^{*}$ is the effective mass of electron.
Note that the distance $a$ is much larger than the lattice constants of the compound materials, making justified the effective mass framework.
Here we use the Fourier representation of the 2D periodic potential with the Fourier coefficients
$V_{\mathbf{G}}$ and the reciprocal lattice vectors $\mathbf{G}=n_1\mathbf{g}_1+n_2\mathbf{g}_2$, with primitive vectors
\begin{align}
    \mathbf{g}_{1(2)}=\frac{2\pi}{3a}(\mathbf{\hat{i}}\pm \mathbf{\hat{j}}\sqrt{3}).
    \label{reciprocal vectors}
\end{align}

The Rashba SOI Hamiltonian is defined as follows
\begin{equation}
    H_{\mathrm{R}}=\frac{\alpha}{\hbar}\left(\boldsymbol{\sigma} \times \left(\mathbf{p}+\frac{e}{c}\mathbf{A}\right)\right)_{z},
    \label{HR}
\end{equation}
where $\mathbf{A}$ is the vector potential of the cavity photon field,
$\alpha$ denotes the Rashba spin-orbit coupling strength,
$\boldsymbol{\sigma} = (\sigma_x, \sigma_y, \sigma_z)$ is the vector of Pauli matrices, $e$ is the elementary charge, $c$ is the speed of light, and $\hbar$ is the reduced Planck's constant.

The electron-photon (e-ph) interaction Hamiltonian contains a paramagnetic term linear in the vector potential and a diamagnetic term quadratic in it, reflecting first- and second-order light-matter coupling. In the Coulomb gauge, it reads as follows
\begin{align}
    H_{\mathrm{e-ph}} 
    &= \frac{e}{m^{*}c}\textbf{A} \cdot \textbf{p} + \frac{e^2}{2m^{*}c^2}A^2.
    \label{H_e-ph}
\end{align}
The cavity field is modeled as a single quantized photon mode with $H_{\mathrm{ph}}=\hbar\omega_{\mathrm{ph}} \hat{a}^{\dagger}\hat{a}$, where $\hat{a}^{\dagger}$ and $\hat{a}$ denote the photon creation and annihilation operators.
The explicit form of the Hamiltonian (\ref{H_e-ph}) depends on the cavity geometry through the vector potential $\textbf{A}$.
Here we consider two geometries for the cavity, namely cylindrical with $\textbf{A}_{\text{cyl}}$ and linear with $\textbf{A}_{\text{lin}}$.

In the long-wavelength approximation, the quantized vector potential of the cylindrical cavity photon mode with amplitude $A_{\mathrm{0}}$ is given by \cite{Gudmundsson2025}
\begin{equation}
     \textbf{A}_{\mathrm{cyl}}=\mathbf{\hat{e}_{\varphi}}A_{\mathrm{0}}(\hat{a}^{\dagger}+\hat{a})r,
     \label{A_cyl}
\end{equation}
where $\mathbf{\hat{e}_\varphi} = -\mathbf{\hat{i}} \sin\phi  + \mathbf{\hat{j}} \cos\phi$ is the unit angular vector in the polar system of coordinates.

For a system positioned in the middle ($z=0$) of a linear cavity with two mirrors fixed at $z=\pm{L/2}$, and confined within a narrow enough width (for a photon state $n$) $d \ll  L/n$ the quantized vector potential for a photon mode with polarization angle $\theta$ with regard to $x$ axis is as follows \cite{Gerry}
\begin{equation}
\textbf{A}_{\mathrm{lin}} = A'_\mathrm{0} \sqrt{\frac{2}{L}}(\hat{a}^\dagger + \hat{a})\! 
\left(\mathbf{\hat{i}} \! \cos\theta + \mathbf{\hat{j}} \! \sin\theta \right).
\label{A_lin}
\end{equation}

It can be easily shown that the diamagnetic and the paramagnetic terms of the e-ph interaction Hamiltonian are proportional to $g$ and $g^2$, respectively, where
\begin{equation}
g_{cyl}=g=\frac{e  a_\mathrm{0}^2}{c\hbar}A_\mathrm{0},\quad g_{lin}=g'=\frac{e  a_\mathrm{0}}{ c\hbar } \sqrt{\frac{2}{L}}A'_0
\end{equation}
are dimensionless coupling constants which naturally emerge from the minimal-coupling substitution
$\mathbf{p}\rightarrow\mathbf{p}+(e/c) \mathbf{A}$,
and $a_0$ stands for the effective Bohr radius of the host material (see Appendix \ref{Real Hamiltonian}).
Further we will consider only $\mathbf{\hat{x}}$ ($\theta = 0$) and $\mathbf{\hat{y}}$ ($\theta = \pi/2$) polarizations for linear cavity.

The single-particle Schr\"odinger equation with light-matter and Rashba interactions leads to coupled spin-component equations which can be written compactly in a matrix form
\begin{equation}
\begin{pmatrix}
H_{11} & H_{12} \\
H_{21} & H_{22}
\end{pmatrix}
\begin{pmatrix}
\Psi_{\uparrow} \\
\Psi_{\downarrow}
\end{pmatrix}
=
E
\begin{pmatrix}
\Psi_{\uparrow} \\
\Psi_{\downarrow}
\end{pmatrix},
\label{Equation}
\end{equation}
where the blocks of the Hamiltonian $H_{12}$ and $H_{21}$ implicitly depend on the chosen cavity type (see Appendix \ref{Real Hamiltonian}).
Using the periodicity of the potential (\ref{Vh}) the spinor components in Eq. (\ref{Equation}) for a state $|\delta\rangle$ can be expanded in a tensor product basis of plane-waves and the photon Fock states
\begin{equation}
\Psi^{(\delta)}_{\textbf{G},s,n}(\textbf{r},\textbf{k}) =
\mathcal{A}^{-1/2}
U^{(\delta)}_{\textbf{G},s,n}(\textbf{k})
e^{i(\textbf{k}+\textbf{G})\cdot\textbf{r}}
\otimes |n\rangle,
\label{basis}
\end{equation}
where, $\mathcal{A}$ is the area of the superlattice unit cell, $\textbf{k}$ is the electron wave vector, $s$ is the spin index, and $|n\rangle$ denotes the photon Fock state. The energy dispersions and the eigenvector coefficients $U^{(\delta)}_{\textbf{G},s,n}(\textbf{k})$ are obtained by exact diagonalization of the system Hamiltonian.

The occupation of the minibands by the cavity mode is characterized by the mean photon number in each of them, defined as the expectation value of the photon number operator
$\langle \Psi_{\alpha} | \hat{a}^{\dagger} \hat{a} | \Psi_{\alpha} \rangle$. 
For a given miniband $\delta$ this reduces to the integration over the FBZ
\begin{equation}
\langle \hat{N}_{\mathrm{ph}} \rangle _{\delta}
=
\frac{\mathcal{A}}{(2\pi)^2}\sum_{\mathbf{G},s,n} 
\int_{\text{FBZ}} d \textbf{k} 
n (\textbf{k}) \, \big| U_{\mathbf{G},s,n}^{(\delta)}(\textbf{k}) \big|^2.
\label{N_ph}
\end{equation}

A finite value of $\langle N_{\mathrm{ph}} \rangle_{\delta}$ indicates the formation of hybrid light-matter states due to e-ph coupling, while $\langle N_{\mathrm{ph}} \rangle_{\delta} \approx 0$ corresponds to a photon vacuum state.

The spin conductivity tensor can be evaluated within the linear-response theory using the Kubo formalism \cite{Marder}
\begin{flalign}
&\sigma_{xy}
=
\frac{-i\hbar}{2(2 \pi)^{2}}
\sum_{\substack{\delta, \beta \\ \delta \neq \beta}} \int_{\text{FBZ}} d \textbf{k} 
\Omega_{s}^\delta(\textbf{k}){f(E(\delta, \textbf{k})) - f(E(\beta,\textbf{k}))}  ,
\label{spin-cond}
\end{flalign}
where, $E(\delta,\mathbf{k})$ and $|\delta, \mathbf{k}\rangle$ are the eigenvalues and eigenstates of the full Hamiltonian, $f(\varepsilon)$ is the Fermi-Dirac distribution function and spin-Berry curvature $\Omega_{s}^\delta(\mathbf{k})$ is as follows \cite{Marder}
\begin{align}
\Omega_{s} ^{\delta}(\mathbf{k}) = -2 \, \text{Im} \sum_{\beta \neq \delta} \frac{\langle \delta, \mathbf{k} | \hat{J}^{z}_x | \beta, \mathbf{k} \rangle \langle \beta, \mathbf{k} | v_y | \delta, \mathbf{k} \rangle}{(E(\delta,\mathbf{k}) - E(\beta,\mathbf{k}))^2}.
\label{spin-BC}
\end{align}
The Berry curvature (BC) is defined as \cite{Xiao}
\begin{align}
\Omega ^{\delta}(\mathbf{k}) = -2 \, \text{Im} \sum_{\beta \neq \delta} \frac{\langle \delta, \mathbf{k} | v_x | \beta, \mathbf{k} \rangle \langle \beta, \mathbf{k} | v_y | \delta, \mathbf{k} \rangle}{(E(\delta,\mathbf{k}) - E(\beta,\mathbf{k}))^2}.
\label{BC}
\end{align}
The non-zero elements of the spin-Hall conductivity tensor are non-diagonal ones. For our system they are evaluated using the expressions $\hat{J}^{z}_{x(y)} = (1/2)\left\{\hat{v}_{x(y)}, s_{z} \right\}$, and $\hat{J}^{e}_{y(x)}=-e\hat{v}_{y(x)}$, where $s_{z}=(\hbar/2) \sigma_{z}$ in the anticomutator. The general definition of the total velocity operator $\textbf{v}_\xi$ is given by:
\begin{equation}
    \textbf{v}_{\xi} = \textbf{v}_{\xi}^{(T)} + \textbf{v}_{\xi}^{(R)} + \textbf{v}_{\xi}^{(cav)},
    \label{velocity}
\end{equation}
where $\textbf{v}_{\xi}^{(T)}$ is the kinetic term, $\textbf{v}_{\xi}^{(R)}$ is the Rashba term, and $\textbf{v}_{\xi}^{(cav)}$ represents the cavity interaction term. The kinetic and Rashba terms are
\begin{align}
    \textbf{v}_{\xi}^{(T)} &= \frac{\mathbf{p}_{\xi}}{m}, \\
    \textbf{v}_{\xi}^{(R)} &= -\frac{\alpha}{\hbar}\left(\sigma_y\hat{\mathbf{i}} - \sigma_x\hat{\mathbf{j}}\right).
\end{align}
The cavity interaction term for a cylindrical cavity is
\begin{equation}
    \textbf{v}_{\text{cyl}}^{(cav)} = -\frac{\hbar}{m a^2} g_{\text{cyl}} (\hat{a}^\dagger + \hat{a})\left(y\hat{\mathbf{i}} - x\hat{\mathbf{j}}\right),
\end{equation}
whereas for the linearly $x$- and $y$-polarized cavities it takes the form
\begin{equation}
   \textbf{v}_{x(y)}^{(cav)} = \frac{\hbar}{m a} g_{\text{lin}} (\hat{a}^\dagger + \hat{a}) 
    \begin{cases} 
        \hat{\mathbf{i}} & \text{for } x\text{-polarization}, \\ 
        \hat{\mathbf{j}} & \text{for } y\text{-polarization}. 
    \end{cases}
\end{equation}
\section{Results and Discussion}
\begin{figure}
\centering
\includegraphics[width=0.7\linewidth]{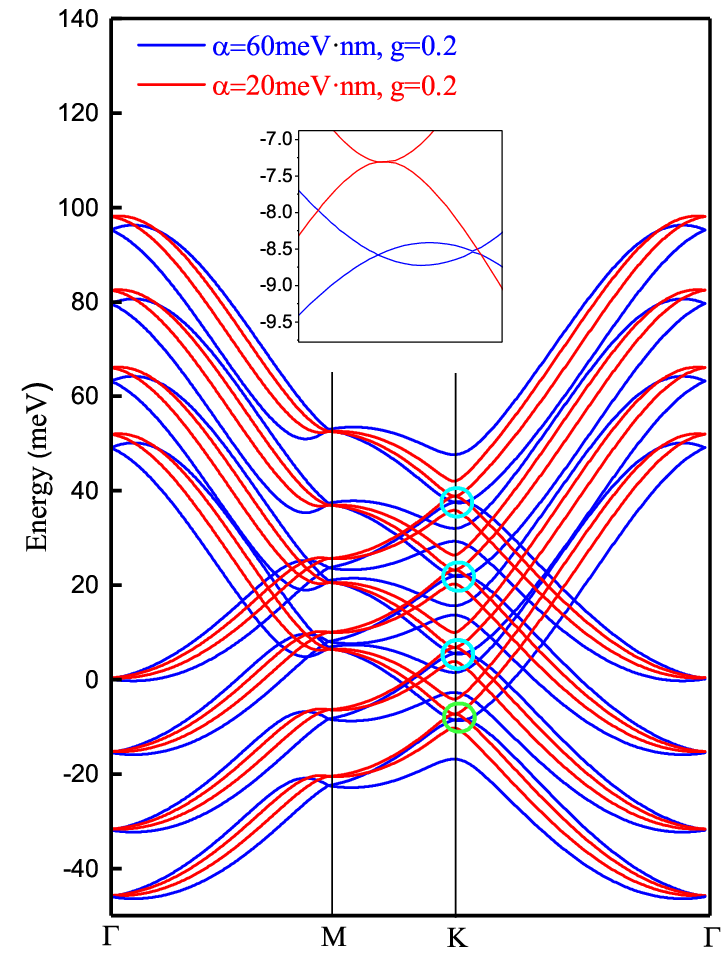}
\caption{Energy miniband dispersion along the chosen k-space path for $\alpha = 60$ and $20$ meV$\cdot$nm, and for interaction constant $g=0.2$. The inset corresponds to the region indicated by a green circle.}
\label{ZoomBandstructure}
\end{figure}
The numerical calculations are carried out for the following values of parameters: $m^{*}=0.023m_{0}$, where $m_{0}$ is the free electron mass, $V_{0}=67 $ meV, $a=12$ nm, and $H_{\mathrm{ph}}=10 $ meV. In the Fermi-Dirac distribution functions in Eq. (\ref{spin-cond}) we take $k_B T = 1.08$meV. We also introduce very small yet non-zero broadening parameters $\lambda = 0.05$meV and $\eta = 0.01$meV in the denominators of Eqs. (\ref{spin-BC}) and (\ref{BC}), respectively.

In Fig.\ \ref{Bandstructure} the band structures for the four replicas of the two lowest minibands of AG in cylindrical ((b), (c), (g), (h)) and linear ((d), (e), (i), (j)) cavities are presented along the path shown in the inset of Fig.\ \ref{Bandstructure} (a). Fig.\ \ref{Bandstructure} (a) and (f) correspond to a non-interacting e-ph system.
In the absence of Rashba SOI (Fig.\ \ref{Bandstructure} (a)-(e)) each miniband is spin-degenerated.
For the system with no interaction or with interaction with photons of the cylindrical cavity all the electronic DPs preserve. The cylindrical cavity does not open a gap in the DP, because it does not affect the triangular symmetry of AG (Figs.\ \ref{Bandstructure} (a)-(c)).
However, the multiplication of minibands leads to many crossings between the replicas. 
\begin{figure}
\centering
\includegraphics[width=1\linewidth]{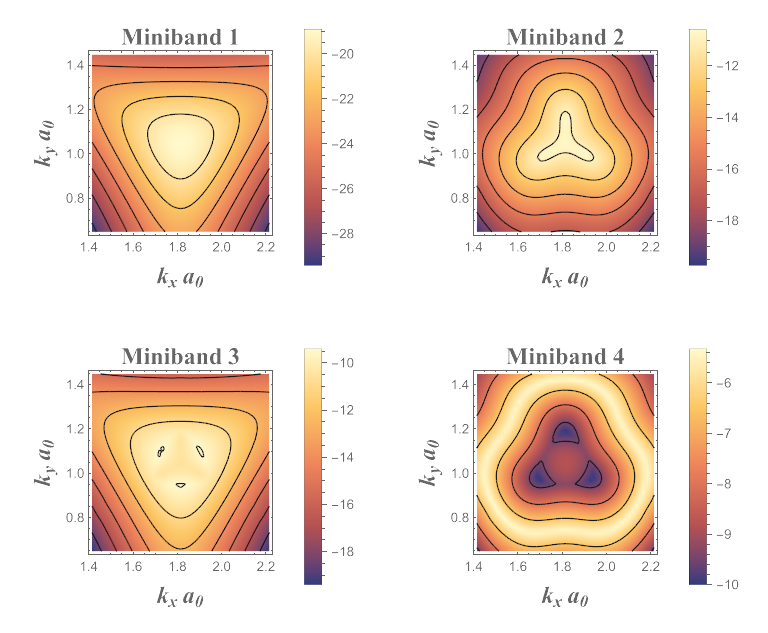}
\caption{Density plots of the lowest four energy minibands in the vicinity of {\lq}{\lq}K{\rq}{\rq} point for $\alpha=60$  meV$\cdot$nm and $g=0.001$.}
\label{DensPlot_g0}
\end{figure}
\begin{figure}
\centering
\includegraphics[width=1\linewidth]{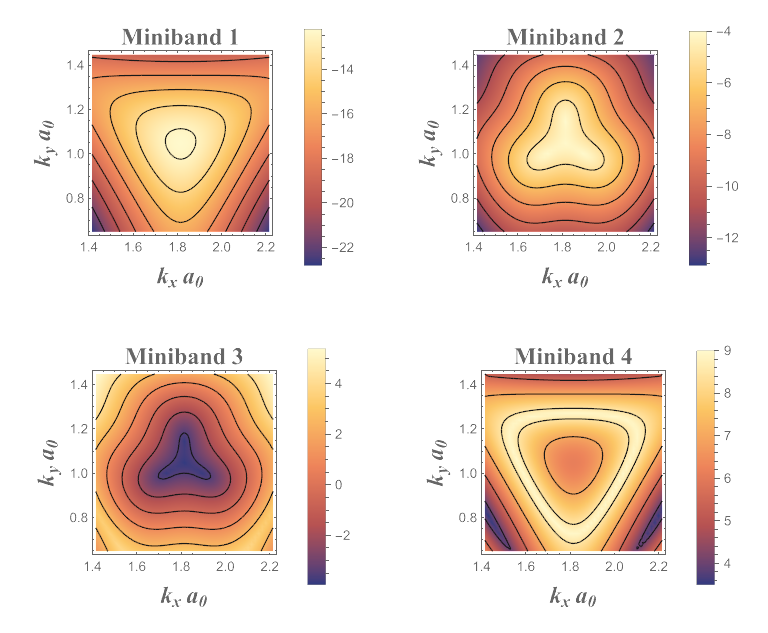}
\caption{Density plots of the lowest four energy minibands for the system inside cylindrical cavity in the vicinity of {\lq}{\lq}K{\rq}{\rq} point for $\alpha=60 $ meV$\cdot$nm and $g=0.4$.}
\label{DensPlot_g04}
\end{figure}
\begin{figure}
\centering
\includegraphics[width=1\linewidth]{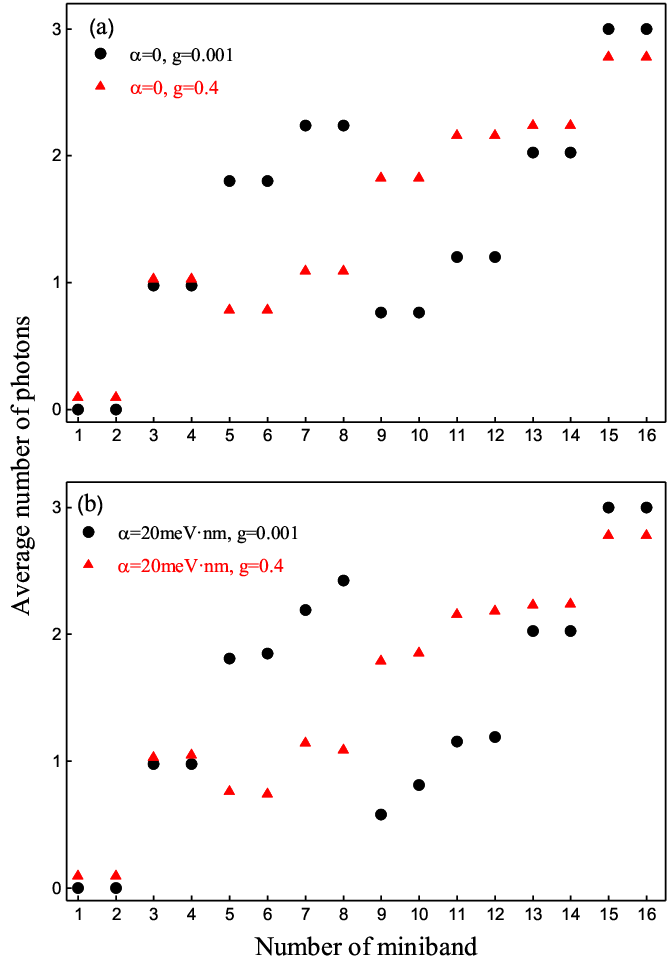}
\caption{Mean photon number per unit cell in each miniband for the AG inside a cylindrical cavity.}
\label{PhNum_cyl}
\end{figure}
\begin{figure}
\centering
\includegraphics[width=1\linewidth]{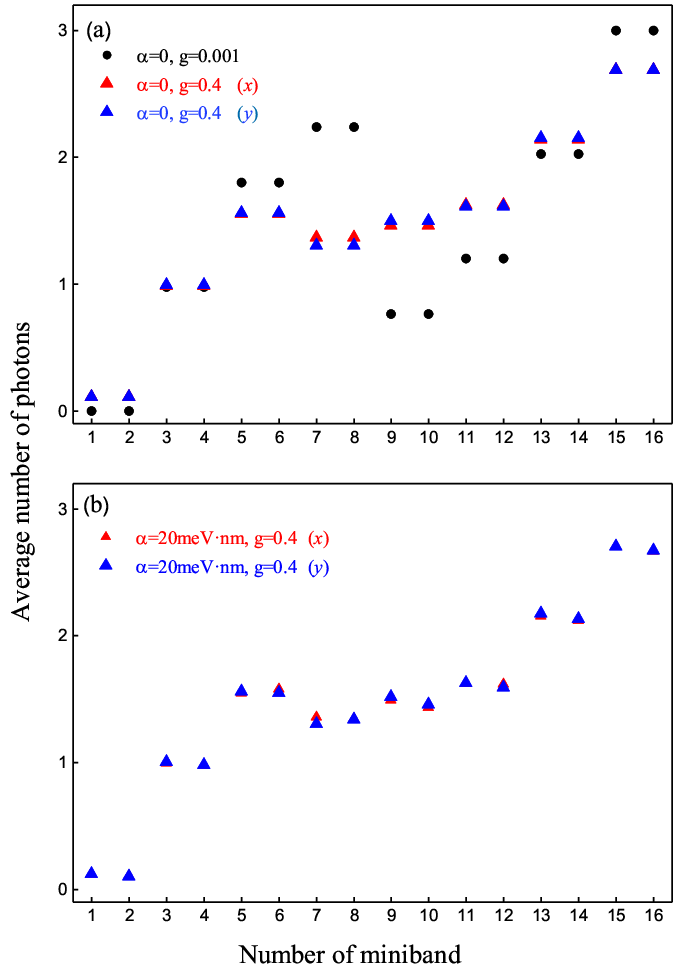}
\caption{Mean photon number per unit cell in each miniband for the AG inside a linear cavity with different polarizations.}
\label{PhNum_lin}
\end{figure}
\begin{figure*}
\centering
\includegraphics[width=1\linewidth]{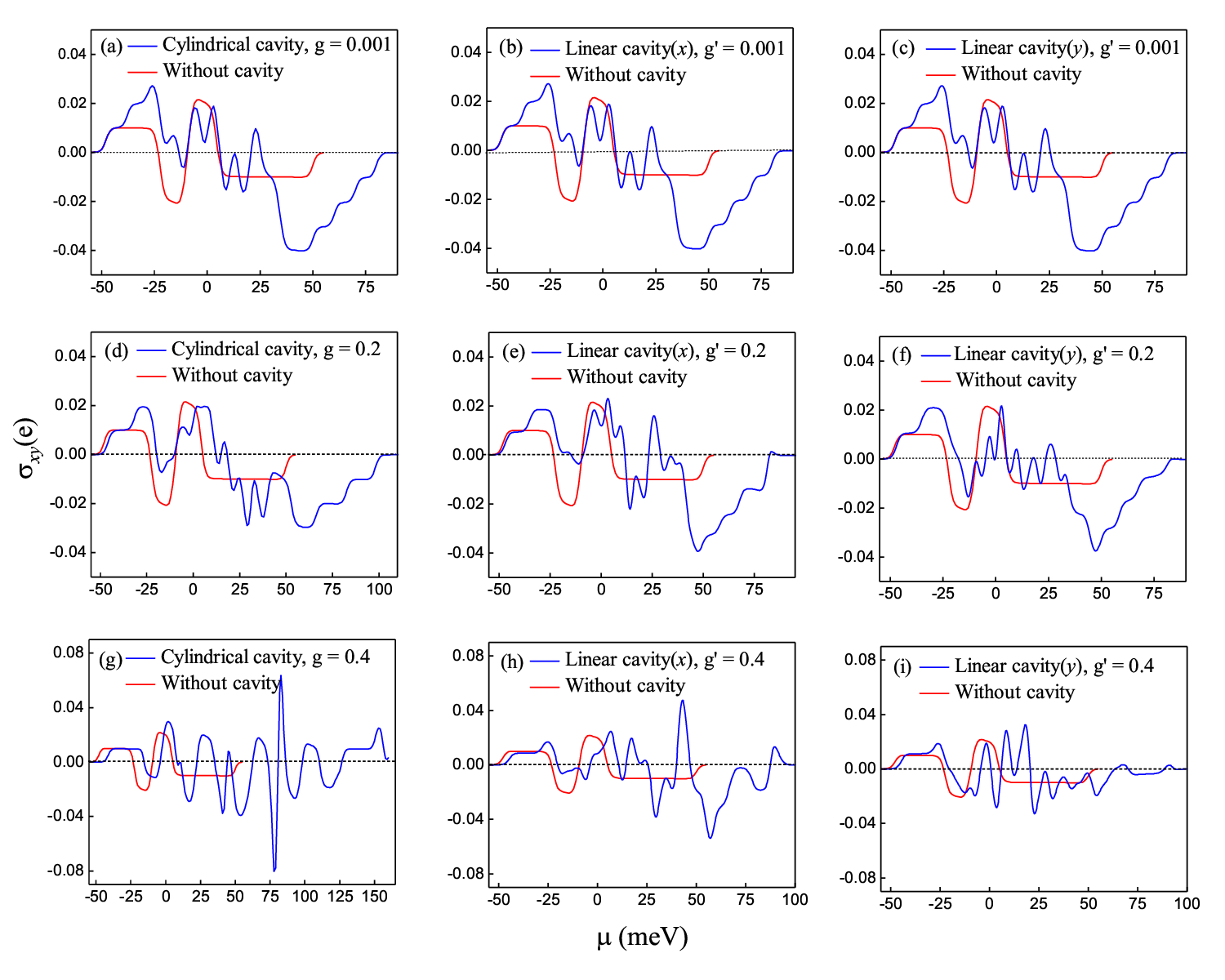}
\caption{Spin-Hall conductivity dependence on the chemical potential for AG inside a cylindrical cavity (the first column), as well as inside an x-polarized (second column) and y-polarized (third column) cavities for different values of e-ph coupling parameter.}
\label{Cond}
\end{figure*}
The comparison of Fig.\ \ref{Bandstructure} (a), (b) and (c) shows that the e-ph interaction has a significant impact on the position and the behavior of the minibands.
Namely, the energy differences between neighboring replicas significantly deviate from the bare photon energy, indicating strong e–ph hybridization and Rabi-type coupling between the minibands.
The Rashba SOI results in the spin-splitting of the minibands (Fig.\ \ref{Bandstructure} (f)-(j)) as was expected.
In Fig.\ \ref{Bandstructure} (f) the touching of two spin-split minibands in the point M (as well as in the center) of the FBZ is due to the vanishing group velocity of electron.
Importantly, these touching points are not shifted in the reciprocal space because the cavity photons (which correspond to standing waves), regardless of the cavity symmetry, do not contribute in the group velocity of polariton (Figs.\ \ref{Bandstructure} (g)-(j)).
Other touchings of the neghboring minibands belonging to different pairs of Rashba-split minibands for the system inside cylindrical cavity are observed in the vicinity of {\lq}{\lq}K{\rq}{\rq} point (touchings of the red and blue lines in Figs.\ \ref{Bandstructure} (f)-(h)).

For a more detailed examination we have calculated the band structure of e-ph system inside a cylindrical cavity for a bigger value of Rashba parameter: $\alpha = 60$meV$\cdot$nm (Fig.\ \ref{ZoomBandstructure}).
Besides the significant increase of the miniband splitting (compare the blue and the red curves in Fig.\ \ref{ZoomBandstructure}) one can observe an additional touching (Dirac) point which is shifted from the {\lq}{\lq}K{\rq}{\rq} point to the center of the FBZ along its diagonal for each split photon replica.
The regions where the additional DPs are observed are indicated by circles on the graph and one of them is also presented as an inset (the region indicated by a green circle).
This result is in complement with previous works where the emergence of three additional DPs around each of \lq \lq K\rq \rq or \lq \lq K$'$\rq \rq points is predicted both in the frame of tight-binding model \cite{Zarea2009}, and exact diagonalization scheme \cite{Mughnetsyan2019PRB}.

Going back to Fig.\ \ref{Bandstructure}, let us discuss the band structure of e-ph system in linear cavities with $x$ -  ((d), (i)) and $y$ -  ((e), (j)) polarizations.
It is obvious that the broken symmetry of the system due to a cavity leads to dramatic changes in the behavior of minibands.
The original degeneracies at DPs are removed (see the lowest and uppermost pairs of photon replicas in Fig.\ \ref{Bandstructure} (d)) or replaced by touching points which are located between the {\lq}{\lq}M{\rq}{\rq} and the {\lq}{\lq}K{\rq}{\rq} points of the FBZ (Figs.\ \ref{Bandstructure} (d) and (e)).
Comparison of Figs.\ \ref{Bandstructure} (d) and (e) shows that these touching points are conventional, or so called type-I DPs (indicated by circles) for the $y$ - polarized cavity, while they are type-II DPs or tilted DPs (indicated by rectangles) for the $x$ - polarized cavity \cite{Mann2018}.

The physical difference between tipe-I and type-II DPs is unveiled by considering Rashba SOI for the system inside a linear cavity (Figs.\ \ref{Bandstructure} (i) and (j)).
Namely, one can observe the appearance of energy gaps between the split minibands due to Rashba SOI for type-II DPs (indicated by rectangles in Fig.\ \ref{Bandstructure} (i) and (j)) because of the anisotropic Fermi velocity around those points, while for the type-I DPs there are still touchings between the neighboring minibands which belong to different Rashba-split pairs (indicated by circles in Fig.\ \ref{Bandstructure} (j), see also Appendix \ref{GapOpening}).

Figs.\ \ref{DensPlot_g0} and \ref{DensPlot_g04} represent the density plots for the dispersion surfaces of the lowest four energy minibands for the Rashba-interacting system near the {\lq}{\lq}K{\rq}{\rq} point in the absence (Fig.\ \ref{DensPlot_g0}) and the presence (Fig.\ \ \ref{DensPlot_g04}) of e-ph interaction in a cylindrical cavity.
A clear triangular symmetry indicates directly on the non-isotropic Fermi velocity both near the {\lq}{\lq}K{\rq}{\rq} point and additional DPs.
The comparison of Figs.\ \ref{DensPlot_g0} and \ \ref{DensPlot_g04} shows that the interaction with cavity photons leads to the qualitative change in the behavior of energy when moving further from the {\lq}{\lq}K{\rq}{\rq} point.
Namely, in the absence of interaction, the gradient of the touching surfaces for the 2-nd and the 3-rd minibands has the same sign (negative for both of them) not far from the {\lq}{\lq}K{\rq}{\rq} point, but when going further the gradient of the 3-rd miniband becomes positive. In contrast, in the presence of the e-ph interaction the gradient of the 3-rd miniband becomes positive for the points outside the triangle composed by the three additional DPs.

The number of photons $\langle N_{\mathrm{ph}} \rangle$ in each miniband averaged over the FBZ without (a) and with (b) Rashba SOI is presented in Figs.\ \ref{PhNum_cyl} and \ \ref{PhNum_lin} for cylindrical and linear cavities, respectively.
It is obvious that without SOI $\langle N_{\mathrm{ph}} \rangle$ is the same for spin-degenerated minibands even in the presence of e-ph interaction (Fig.\ \ref{PhNum_cyl} (a) and Fig.\ \ref{PhNum_lin} (a)).
The mean number of photons is a non-monotonic function on the number of miniband. 
The divergence of $\langle N_{\mathrm{ph}} \rangle$ from integer numbers for the minibands different from the lowest (1st and 2nd) and highest (15-th and 16-th) ones is a result of crossings between the different photon replicas.

In a cylindrical cavity (Fig.\ \ref{PhNum_cyl}) the e-ph interaction for the first and the second pair of minibands leads to a slight increase in $\langle N_{\mathrm{ph}} \rangle$, while for the highest couple of minibands (15-th and 16-th) the number of photons decreases.
The strongest effect the e-ph interaction has on the intermediate minibands which are closer to each other and reveal many crossings.
Namely, the significant decrease of the mean-photon number in the 3-rd and the 4-th pairs of minibands is compensated by a drastic increase of $\langle N_{\mathrm{ph}} \rangle$ in the 5-th, 6-th and the 7-th pairs of the minibands.
\begin{figure}
\centering
\includegraphics[width=1\linewidth]{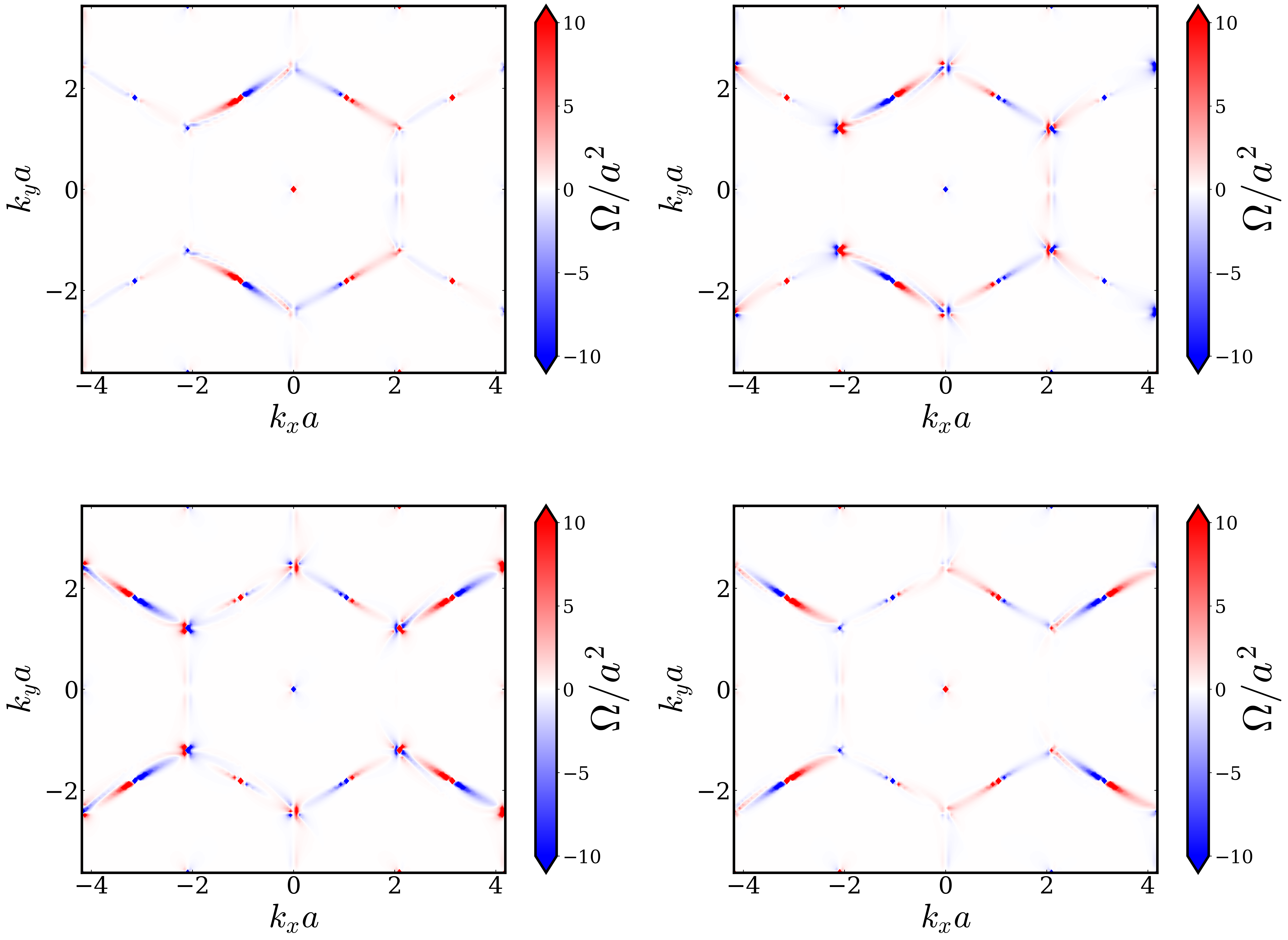}
\caption{The distribution of BC for the lowest minibands in AG without cavity for $\alpha = 20$ meV$\cdot$nm.}
\label{BC_e}
\end{figure}
\begin{figure}
\centering
\includegraphics[width=1\linewidth]{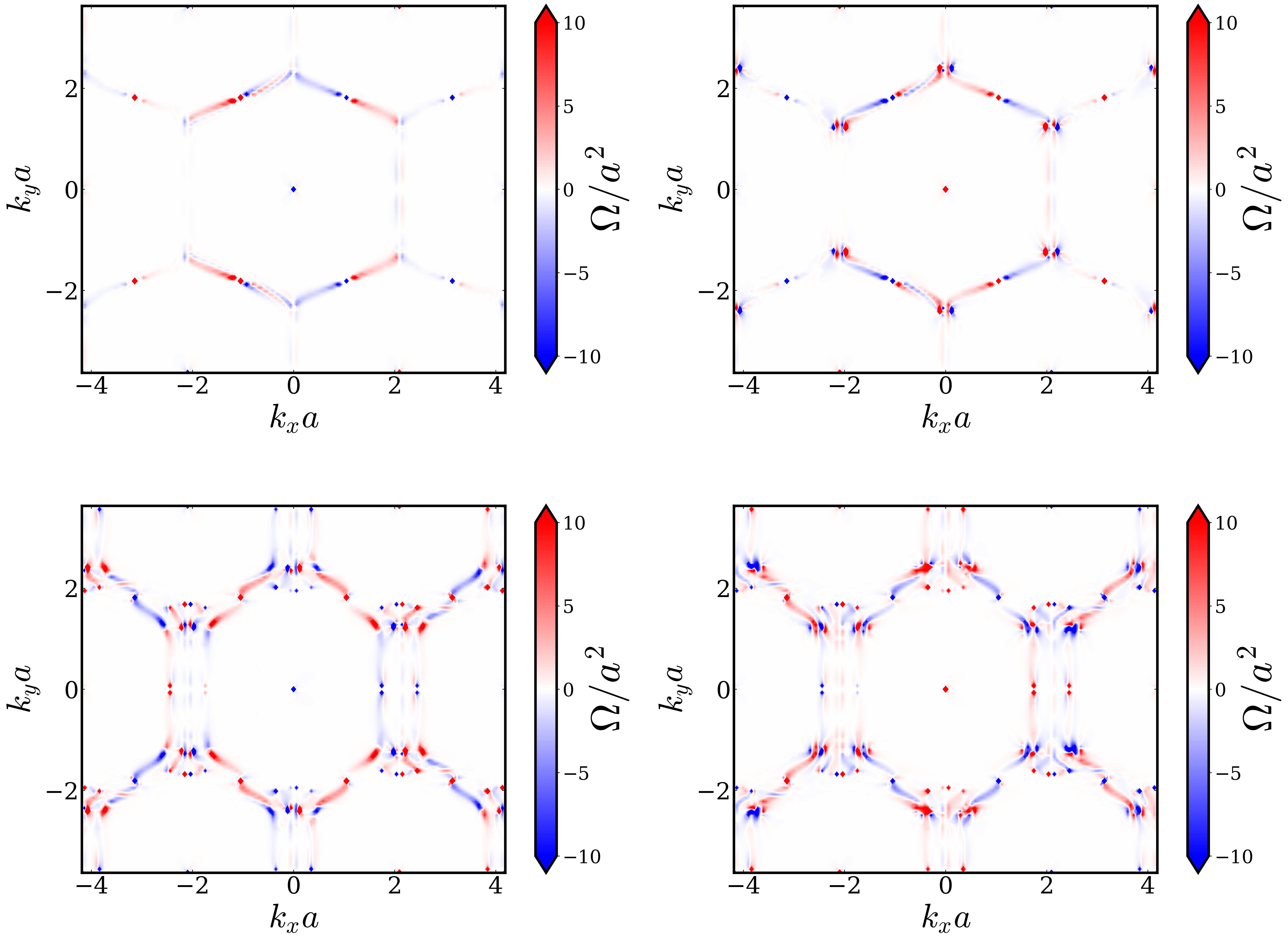}
\caption{The distribution of BC for the lowest minibands in AG with x-polarized cavity for $\alpha = 20$ meV$\cdot$nm, and $g' = 0.4$.}
\label{BC_x}
\end{figure}
\begin{figure}
\centering
\includegraphics[width=1\linewidth]{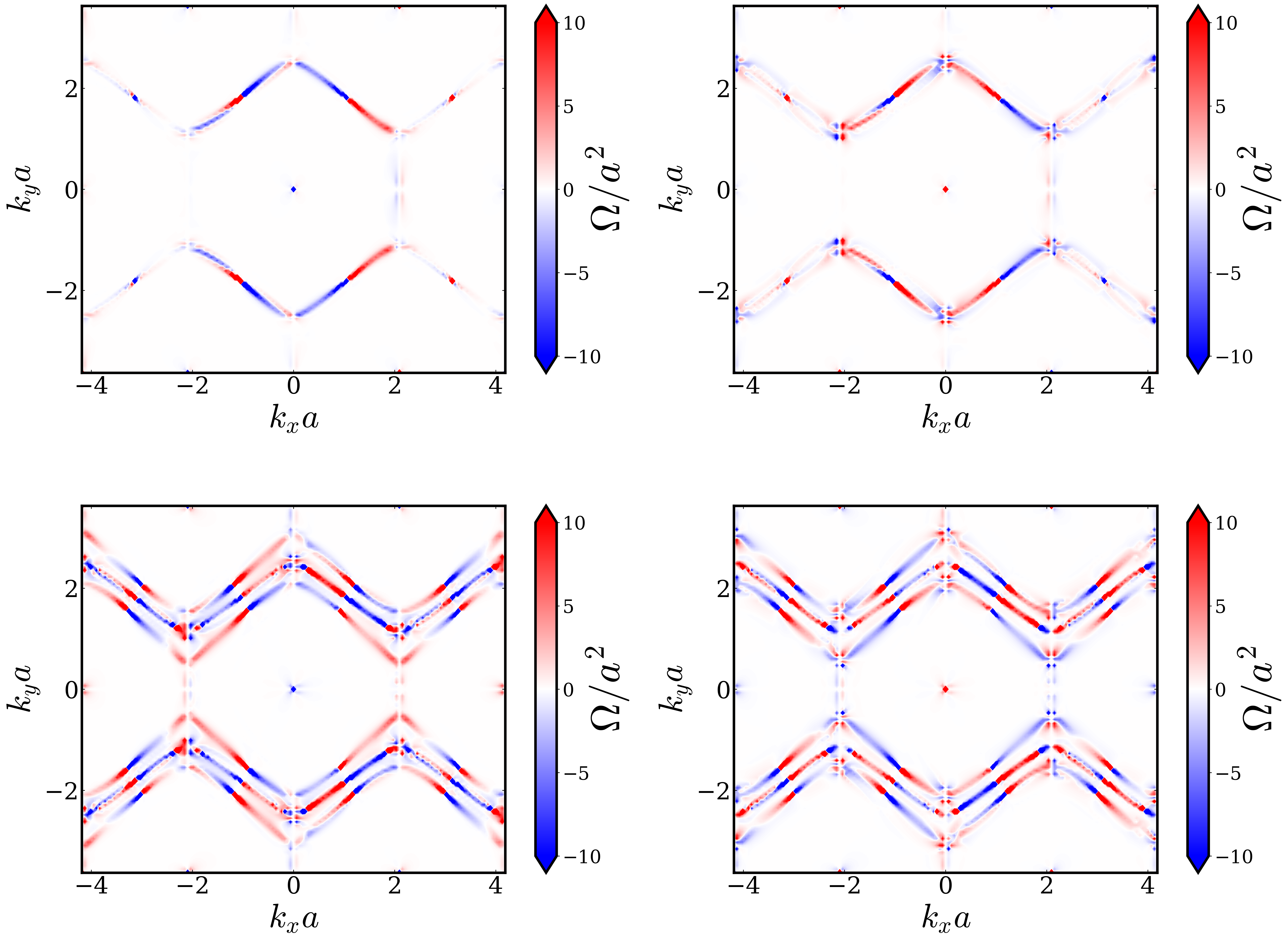}
\caption{The distribution of BC for the lowest minibands in AG with y-polarized cavity for $\alpha = 20$ meV$\cdot$nm, and $g' = 0.4$.}
\label{BC_y}
\end{figure}
The consideration of Fig.~6(b) shows that the degeneracy
with regard to $\langle N_{\mathrm{ph}} \rangle$ is lifted along with the degeneracy
with regard to spin as a result of Rashba SOI.
Interestingly, for the 3-rd and the 4-th pairs of minibands
the increase of $\langle N_{\mathrm{ph}} \rangle$ is replaced by its decrease
due to e-ph interaction, while for other minibands this is not the case.

Comparing Fig.\ \ref{PhNum_cyl} and Fig.\ \ref{PhNum_lin} one can conclude that the effect of the linear cavity on the mean photon number is similar with one of the cylindrical cavity. The effect of linear cavity is slightly stronger for the lowest and the uppermost miniband pairs and is weaker for the rest of minibands (Fig.\ \ref{PhNum_lin} (a)). The $x$ - and $y$ - polarized photons have almost the same effect on the number of photons regardless the Rashba SOI.

Fig.\ \ref{Cond} demonstrates the  spin-Hall conductivity as a function of the chemical potential for AG embedded in cylindrical and linear FIR cavities at different e-ph coupling strengths. The red curves correspond to the purely electronic system, while the blue curves represent the cavity-coupled polaritonic states. The calculations are performed for the parameter of the Rashba SOI $\alpha=20~\mathrm{meV \cdot nm}$, where spin splitting and nontrivial Dirac-point physics are present.
For very small values of the coupling parameter ($g,g'=0.001$) the conductivity exhibits relatively smooth variations with the chemical potential and its behavior is obviously the same for all the polarizations of the cavity. 
These features originate mainly from Rashba-induced spin splitting of the minibands and crossing of different photon replicas hosting type-I Dirac points preserved by the electronic structure.
The step-like behavior of the conductivity in the presence of weakly-interacting photon field is the result of involving the contribution of new replicas with increasing chemical potential.

The inclusion of the cavity field qualitatively changes the transport response of the system. Even for moderate coupling ($g,g'=0.2$), pronounced oscillatory structures emerge in the conductivity spectra. These oscillations are a direct manifestation of the hybridization between electronic minibands and photon replicas. Since the e-ph interaction generates additional anticrossings in the case of linear cavity and modifies the local miniband gradients, the corresponding BC becomes significantly redistributed in momentum space, producing large conductivity fluctuations.

At strong e-ph coupling ($g,g'=0.4$), the hybridization effects lead to substantial modifications and
pronounced oscillatory behavior of the conductivity. The cylindrical cavity exhibits the densest sequence of oscillations over the considered chemical-potential range, reflecting the large number of crossings between photon-dressed minibands. The spin-Hall conductivity in the $x$-polarized linear cavity develops a pronounced oscillatory structure with several asymmetric peaks and multiple sign reversals as the chemical potential is varied. Compared with the weak-coupling regime, the conductivity becomes significantly more sensitive to the position of the Fermi level, indicating a substantial reconstruction of the miniband spectrum induced by the interplay between e-ph coupling and Rashba SOI. The appearance of additional crossings and anticrossings between photon-dressed minibands leads to a redistribution of the BC in momentum space (see Fig.\ref{BC_x}), which manifests itself in the observed conductivity fluctuations.

The $y$-polarized linear cavity also exhibits oscillatory behavior and several sign reversals of the conductivity, demonstrating that the cavity field strongly reconstructs the miniband structure in this case as well. However, the oscillations are concentrated within a narrow range of chemical potential and the conductivity is substantially suppressed away from this region. In comparison with the $x$-polarized cavity, the resonant features are less extended over energy and the conductivity approaches zero more rapidly at larger chemical potentials. The distinct responses of the two polarizations highlight the pronounced anisotropy introduced by the cavity field.

Figure \ref{BC_e} shows that in the absence of the cavity the BC is predominantly localized near the multiplied Dirac touchings induced by the Rashba SOI and concentrated along the boundaries of the FBZ. We would like to emphasise that the time reversal symmetry of the system is preserved (which leads to the total BC and the Chern number of the system being 0), nevertheless the BC does not exhibit an antisymmetric behavior in a single band. This fact is connected with the Rashba splitting between the minibands due to which the time reversal is note simply connected with the sign-change of the quasimomentum but is also associated with \lq \lq jumps\rq \rq between the minibands. The peaks of BC at the center and in the middles of the edges of FBZ is due to the touching between the Rashba-coupled minibands. Furthermore, the existance of three positive and three negative peaks which are rotated by $\pi/3$ with regard to each other near the corners of the FBZ for the 2-nd and the 3-rd minibands is connected with the position of closely located additional SOI-generated Dirac points around the original ones. 

The introduction of the cavity field strongly reconstructs the geometric properties of the minibands through the hybridization of electronic states with photon replicas. For the $x$-polarized cavity (Fig.\ref{BC_x}) a significant deformation of the BC distribution is observed. For the higher minibands (the second row of the figure) BC redistributes into extended filament-like structures spanning large portions of the FBZ and becomes concentrated around the newly formed cavity-induced avoided crossings. This behavior reflects the strong modification of the underlying miniband topology by photon dressing.

The $y$-polarized cavity (Fig.\ref{BC_y}) produces a qualitatively different redistribution of the BC. Instead of the vertical channels observed for the $x$ polarization, the curvature develops multiple nearly parallel branches following the anisotropically distorted miniband contours. The resulting BC texture directly demonstrates the strong polarization dependence of the e-ph interaction.

It should be emphasized that in the cavity-coupled system the minibands are indexed according to their instantaneous energy ordering, while numerous crossings and anticrossings occur between different photon replicas, as illustrated in Fig.\ref{Bandstructure}. Consequently, the two minibands originating from a given Rashba pair cannot always be uniquely identified throughout the entire FBZ. As a result, the sum of the Chern numbers associated with individual Rashba-split miniband pairs is generally not exactly zero due to the redistribution of BC among different photon-dressed replicas. Nevertheless, time-reversal symmetry remains preserved and enforces an exact cancellation of the total topological charge when all minibands are included, yielding a vanishing sum of Chern numbers for the complete spectrum considered in the calculations.

\section{Conclusion}
We investigate the combined influence of a far-infrared cavity field and Rashba spin–orbit interaction on the band structure and transport properties of artificial graphene formed by quasi-2D InAs/GaAs quantum dots. The light–matter coupling is treated within a composite basis constructed from the electronic Hilbert space and the photon Fock space. For systems embedded in a linear cavity, both type-I and type-II Dirac points emerge and can be clearly distinguished by their response to Rashba interaction: while type-I points remain gapless, a gap opens at type-II Dirac points. The gap induced at type-II Dirac points, on the order of a few meV, originates from the cooperative action of spin–orbit and cavity couplings.

For both linear and cylindrical cavities, we find the formation of electron–photon hybrid states and Rabi splittings between minibands. The resulting band replicas exhibit crossings, and anticrossings, leading to pronounced modifications of the conductivity. In particular, the cavity-induced reconstruction of the miniband spectrum produces strong oscillatory and anisotropic behavior of the spin-Hall conductivity, especially in the presence of linearly polarized cavity modes. The conductivity response becomes increasingly sensitive to the chemical potential with increasing electron–photon coupling strength, reflecting the formation of hybrid polaritonic minibands and the redistribution of the Berry curvature in the Brillouin zone.

Our results demonstrate that the interplay between cavity photons and Rashba spin–orbit interaction provides an efficient mechanism for engineering the band topology and transport properties of artificial graphene. The ability to manipulate Dirac-point geometry, miniband hybridization, and conductivity anisotropy by external cavity fields and tunable spin-orbit interaction opens promising perspectives for controllable polaritonic and topological quantum devices.

\section{Acknowledgments}
This work was financially supported by the Armenian State Committee of Science (grants No 24LCG-1C004 and No 24WS-1C040), by the Research Fund of the University of Iceland Grant No. 92199, and the Icelandic Infrastructure Fund for \lq \lq Icelandic Research e-Infrastructure (IREI)\rq \rq. The computations were performed on resources of the Center for Modeling and Simulations of Nanostructures (NanoSM) at Yerevan State University.
\appendix
\section{Hamiltonian Matrix Elements}
\label{Real Hamiltonian}
The single-particle Hamiltonian describing electrons in a two-dimensional hexagonal lattice with Rashba SOI and a quantized cavity field can be expressed in spin space as
\begin{equation}
\mathcal{H} =
\begin{pmatrix}
H_{11} & H_{12} \\
H_{21} & H_{22}
\end{pmatrix}.
\end{equation}
The diagonal terms include kinetic energy, lattice potential, and photon energies, as well as light-matter coupling energy, while the off-diagonal terms describe Rashba SOI which is modified by the cavity photons:
\begin{equation}
\begin{aligned}
H_{11} &= H_{22}
= -\frac{\hbar^2}{2 m^*} \nabla^{2}
+ \sum_{\mathbf{G}} V_{\mathbf{G}} e^{i \mathbf{G} \cdot \mathbf{r}} \\
&\quad + H_{\text{e-ph}}
+ \hbar \omega\, \hat{a}^\dagger \hat{a}, \\[1mm]
H_{12} &= \alpha (\partial_x - i \partial_y)
+ \frac{e \alpha}{c \hbar} \Big( A_y+iA_x \Big), \\
H_{21} &= \alpha (-\partial_x - i \partial_y)
+ \frac{e \alpha}{c \hbar} \Big( A_y-iA_x \Big).
\end{aligned}
\end{equation}
The e-ph interaction Hamiltonian ($H_{\text{e-ph}}$) arises from the minimal-coupling substitution and in the Coulomb gauge is expressed by Eq.(\ref{H_e-ph}).
\subsection{Cylindrical Cavity Geometry}

For a cylindrical cavity configuration, substituting Eq.(\ref{A_cyl}) into Eq.(\ref{H_e-ph}), we obtain $H_{\text{e-ph}}$ as follows
\begin{equation}
H_{\text{e-ph}}
= g \frac{\hbar}{m^{*} a^2} (\hat{a} + \hat{a}^{\dagger}) \, \mathbf{r} \cdot \mathbf{p}
+ g^2 \frac{\hbar^2}{2 m^{*} a^4} (\hat{a} + \hat{a}^{\dagger})^2 \, \mathbf{r}^2.
\end{equation}
The contribution of e-ph coupling in the Rashba SOI and the paramagnetic interaction are linearly proportional,
while the diamagnetic contribution is quadratically proportional to $g$.

The matrix elements for the non-diagonal blocks of the Hamiltonian are
\begin{equation}
\begin{aligned}[t]
\langle \mathbf{G}', m | H_{12/21} | \mathbf{G}, n \rangle 
= \mathbf{Q} \, I_1,
\end{aligned}
\end{equation}
where
\begin{equation}
\begin{aligned}[t]
\mathbf{Q} &= \alpha \left[ (k_y + G_y) \pm i (k_x + G_x) \right] a \, \delta_{\mathbf{G} ,\mathbf{G}'} \, \delta_{m,n} \\
&\quad + \alpha g  A_{mn}.
\end{aligned}
\end{equation}
and
\begin{equation}
   A_{mn}=  \sqrt{n+1} \, \delta_{m,n+1} 
+ \sqrt{n} \, \delta_{m,n-1}. 
\end{equation}

The paramagnetic contribution to the matrix element is given by

\begin{equation}
\begin{aligned}[t]
g \frac{\hbar}{m^* a^2} \langle \mathbf{G}', m | (\hat{a} + \hat{a}^\dagger) \, \mathbf{r} \cdot \mathbf{p} | \mathbf{G}, n \rangle 
= g \frac{\hbar}{m^* a^2} A_{mn} I_2.
\end{aligned}
\end{equation}

The diamagnetic contribution to the matrix element takes the form
\begin{equation}
\begin{aligned}[t]
g^2\frac{\hbar^2}{2 m^* a^4} \langle \mathbf{G}', m | (\hat{a}^\dagger + \hat{a})^2 r^2 | \mathbf{G}, n \rangle
= g^2\frac{\hbar^2}{2 m^* a^4} B_{mn} I_3,
\end{aligned}
\end{equation}
where
\begin{equation}
\begin{aligned}
B_{mn} &=
\sqrt{(n+1)(n+2)} \, \delta_{m,n+2} 
 + (2n+1) \, \delta_{m,n} \\
&\quad + \sqrt{n(n-1)} \, \delta_{m,n-2},
\end{aligned}
\end{equation}
and $I_1$, $I_2$ and $I_3$ are defined as follows
\begin{align*}
I_1 &=
\int d\mathbf r\, (x \pm i y)\,
e^{-i(\mathbf G' - \mathbf G)\cdot \mathbf r}, \\[2mm]
I_2 &=
(\mathbf k + \mathbf G)\cdot
\int d\mathbf r\, (\mathbf{\hat{j}} x - \mathbf{\hat{i}} y)\,
e^{-i(\mathbf G' - \mathbf G)\cdot \mathbf r}, \\[2mm]
I_3 &=
\int d\mathbf r\, (x^2 + y^2)\,
e^{-i(\mathbf G' - \mathbf G)\cdot \mathbf r}.
\end{align*}
In Eqs. (A4)-(A9) $m$ and $n$ stay for photon numbers.
\subsection{Linearly Polarized Cavity Field}
For a linearly polarized cavity field at angle $\theta$ relative to $x$ axis, using Eq.(\ref{A_lin}), the interaction term $H_{\text{e-ph}}$ is expressed as follows
\begin{equation}
\begin{aligned}
H_{\text{e-ph}}
= g' \frac{\hbar}{m^{*} a^2}\sqrt{\frac{2}{L}} (\hat{a} + \hat{a}^{\dagger})(\hat{\mathbf{i}} \cos\theta+\hat{\mathbf{j}} \sin\theta) \,  \cdot \mathbf{p}\,\\
+ g'^2 \frac{\hbar^2}{2 m^{*} a^4}(\sqrt{\frac{2}{L}})^2 (\hat{a} + \hat{a}^{\dagger})^2 \, .
\end{aligned}
\end{equation}
The corresponding matrix elements for the non-diagonal blocks of the Hamiltonian
\begin{flalign}
&\langle \mathbf{G}', m | H_{12/21} | \mathbf{G}, n \rangle 
= \mathbf{Q'}, 
\end{flalign}
where $Q'$ is defined as $Q$ in (A5) whit the replacement of $g$ by $g'$.
The paramagnetic  contribution to the e-ph interaction is represented by the following matrix element:
\begin{flalign}
&g' \frac{\hbar}{m^{*} a^2}\sqrt{\frac{2}{L}}\langle \mathbf{G'}, m | (\hat{a}+\hat{a}^\dagger)\Big( \mathbf{\hat{i}} \cos\theta + \mathbf{\hat{j}} \sin\theta \Big) \cdot\mathbf{p} | \mathbf{G}, n \rangle = \nonumber \\
&g' \frac{\hbar}{m^{*} a^2}\sqrt{\frac{2}{L}} \Big( (\mathbf{k} + \mathbf{G}) \cdot \Big( \mathbf{\hat{i}} \cos\theta + \mathbf{\hat{j}} \sin\theta \Big) \Big) \notag A_{mn},
\end{flalign}
and the diamagnetic contribution is defined as follows:
\begin{flalign}
& g'^2 \frac{\hbar^2}{2 m^{*} a^4}\Big(\sqrt{\frac{2}{L}} \Big)^2\langle \mathbf{G'}, m | (\hat{a} + \hat{a}^{\dagger})^2 | \mathbf{G}, n \rangle = \nonumber \\
& g'^2 \frac{\hbar^2}{2 m^{*} a^4}\Big(\sqrt{\frac{2}{L}}\Big)^2 B_{mn}
\end{flalign}.

\section{The Dirac-Rashba Hamiltonian for AG in a linear cavity}
\label{GapOpening}
Near a band touching point the Hamiltonian $H'$ describing Dirac electrons with Rashba SOI and interacting with a cavity photon field can be expressed as follows
\begin{equation}
   H' = H'_0 + V_{\sigma A},
   \label{perturbativeH}
\end{equation}
where
\begin{equation}
    H'_0 = H_{\mathrm{ph}} + H_{\mathrm{D}} + H'_{\mathrm{R}}.
    \label{nonperturbativeH}
\end{equation}
Here $H_\mathrm{D} = v_\mathrm{F} \boldsymbol{\sigma} \cdot \mathbf{q}$ represents the Dirac electrons with the Fermi velocity $v_\mathrm{F}$ and quasimomentum $\mathbf{q}$ (counted from the touching point),
$H'_\mathrm{R} = \lambda_R (\sigma_x s_y - \sigma_y s_x)$ accounts for the Rashba spin-orbit interaction
and the interaction with a linear cavity field is described by
\begin{equation}
    V_{\sigma A} = \frac{ev_F A'_0}{\hbar c} \sqrt{\frac{2}{L}} (\sigma_x \cos\theta+\sigma_y \sin\theta) (\hat{a} + \hat{a}^\dagger).
    \label{perturbationV}
\end{equation}
We will consider the later as a perturbation and calculate its impact on the energy at the touching point.
To eliminate the e-ph coupling term, we perform the unitary transformation
\begin{equation}
    H_{\mathrm{eff}} = e^{\hat{S}} H' e^{-\hat{S}},
    \label{H_eff}
\end{equation}
where an anti-Hermitian generator of the transformation
\begin{equation}
    \hat{S} = \frac{G}{\hbar \omega_\mathrm{ph}} (\hat{a}^\dagger - \hat{a})
    \left( \sigma_x \cos\theta + \sigma_y \sin\theta \right),
    \label{generator}
\end{equation}
with $G=e v_\mathrm{F} A_0 \sqrt{2}/\sqrt{L}$.
Near the touching point ($q \rightarrow 0$) the first-order Baker-Campbell-Hausdorff expansion of the effective Hamiltonian (\ref{H_eff}), taking into account (\ref{generator}), can be reduced to
\begin{equation}
    H^{(1)}_{\mathrm{eff}}=H'_\mathrm{R}+H_\mathrm{ph}+[\hat{S},H'_\mathrm{R}]+[\hat{S},V_{\sigma A}],
    \label{H1_eff}
\end{equation}
where
\begin{equation}
    [\hat{S}, V_{\sigma A}] = -2 \frac{G^2}{\hbar \omega_\mathrm{ph}}
    \label{SV}I 
\end{equation}
and
\begin{align}
    [\hat{S}, H_R] &= \alpha_0 (\cos \theta \cdot \sigma_x \otimes \sigma_z + \sin \theta \cdot \sigma_y \otimes \sigma_z) = \notag \\
    &= \alpha_0 \begin{pmatrix} 0 & 0 & e^{-i\theta} & 0 \\ 0 & 0 & 0 & -e^{-i\theta} \\ e^{i\theta} & 0 & 0 & 0 \\ 0 & -e^{i\theta} & 0 & 0 \end{pmatrix}, 
    \label{SH_R}
\end{align}
with 
\begin{align}
    \alpha_0 = 2i \frac{\lambda_R G}{\hbar \omega_{\mathrm{ph}}} (\hat{a}^\dagger - \hat{a}).
\end{align}
It can be shown that the term (\ref{SH_R}) in the Hamiltonian (\ref{H1_eff}) generates an energy gap which is proportional to $\lambda_R$ and $A'_{0}$ and hence is a consequence of the combined effect of the SOI and cavity effects.
\end{document}